\documentclass[11pt]{article}
\usepackage{epsfig,amscd,amssymb}
\usepackage{amsmath,graphicx}
\newcommand{\be}{\begin{equation}}
\newcommand{\ee}{\end{equation}}
\newcommand{\bea}{\begin{eqnarray}}
\newcommand{\eea}{\end{eqnarray}}

\begin{document}

\title{Critical points in coupled Potts models\\ and critical phases in
  coupled loop models}%
\author{Paul Fendley${}^{1,2}$ and Jesper L. Jacobsen${}^{3,4}$
\medskip \\ 
$^{1}$All Souls College and the Rudolf Peierls Centre for Theoretical
Physics,\\ University of Oxford, 1 Keble Road,  OX1 3NP, UK\\
$^{2}$Department of Physics, University of Virginia,
Charlottesville, VA 22904-4714 USA\\
$^{3}$Universit\'e Pierre et Marie Curie, 4 place Jussieu,
75252 Paris Cedex 05, France \\
$^{4}$ Institut de Physique Th\'eorique, CEA Saclay,
91191 Gif-sur-Yvette, France}

\smallskip 


\maketitle

\begin{abstract} 
We show how to couple two critical $Q$-state Potts models to yield a
new self-dual critical point. We also present strong evidence of a
dense critical phase near this critical point when the Potts models
are defined in their completely packed loop representations. In the
continuum limit, the new critical point is described by an $SU(2)$
coset conformal field theory, while in this limit of the
the critical phase, the two loop models decouple.  Using a combination
of exact results and numerics, we also obtain the phase diagram in the
presence of vacancies.  We generalize these results to coupling two
Potts models at different $Q$.

\end{abstract} 

\section{Introduction}

The $Q$-state Potts model is of fundamental interest in
two-dimensional statistical physics. It arises in an amazing number of
different contexts in mathematics and theoretical physics alike:
algebra, combinatorics, probability theory, graph theory,
integrability, conformal field theory (CFT), stochastic Loewner
evolution (SLE), etc. Thus progress in understanding the Potts model
serves as a useful gauge for progress in many of these fields.

Via its underlying algebraic formulation---in terms of the
Temperley-Lieb algebra \cite{TL}---the Potts model can be recast in
several similar ways, but which do however present subtle
differences. Its formulation as a model of completely packed loops
with non-local weights is prototypical within the large (and still
growing) class of so-called {\em loop models} in classical statistical
mechanics. Such models push the limits of CFT, as they present
non-unitary and logarithmic features \cite{PRZ}, and they link
directly to physics-inspired models of extended objects, such as
polymers. The loops underlying the Potts model role as convenient
lattice regularizations of SLE traces \cite{RC}. They can also be
interpreted as particle trajectories in various models of disordered
systems, such as the spin quantum Hall effect \cite{GLR}.

Studying several {\em coupled Potts models} is a challenge for all of
the techniques listed above, and also serves as a worthwhile purpose
on its own right.

In particular, coupled Potts models arise naturally when studying {\em
quantum} loop models \cite{FF,Fendley08}.  Each configuration in the
classical loop model is a basis element of the Hilbert space of the
quantum loop model.  The reason why two coupled copies of the loop
model arise in this context is elementary quantum mechanics. When
computing expectation values, one of course weights by the
wavefunction squared. Thus one loop model comes from the bra, and the
other from the ket. Our results are very useful for quantum loop
models, because the critical phase we find below in the classical
model implies the existence of a phase with {\em topological order} in
the quantum loop model with completely packed loops
\cite{Freedman01,FF,Fendley08}.  In a topological phase, the
excitations are fractionalized, and have anyonic statistics. These
features are possible when the ground state of the quantum system is a
sum over different loop configurations; the non-locality of loops
makes anyonic braiding possible. For the anyonic excitations to be
deconfined, the underlying classical loop model must be critical, so
that all lengths of loops appear in the partition function.

Coupled Potts models are also at play when studying a single Potts
model with quenched random bond disorder. Using the replica trick,
this corresponds to coupling ${\cal N}$ pure Potts models via the
product of their energy operators, followed by the formal limit ${\cal
  N} \to 0$. For $Q-2 \ll 1$ and in the limit of weak non-frustrated
disorder, this can be done within the perturbative renormalization
group (RG) approach of CFT \cite{CardyLudwig}. There exists ample numerical
evidence that the replica symmetry is not broken in this
case. Moreover, the formal expressions obtained for the critical
exponents corresponding to ${\cal N} > 2$ integer have been
identified---order by order in the perturbation---with those of
particular critical points of ${\cal N}$ coupled lattice models.  If
exact results for these critical points could be obtained
exactly, their analytical continuation to the ${\cal N} \to 0$
limit would provide exact results for the random-bond Potts model
\cite{DJLP}. In this context, the case of two coupled models to be
studied here is somewhat particular, since the RG flows depend on
the coupling $g$ in the combination $(2-{\cal N})g$.

In the present study we assume (as in \cite{DJLP}) that the coupled
Potts models are self-dual, i.e., they are left invariant under the
combined duality transformation of both underlying models.  Although
it is known in the special case of $Q=2$ that mutually dual pairs of
critical points do exist, it seems plausible that self-duality will
generically allow for the highest multicritical---and thus most
interesting---behavior.

Among the directions that we leave for future investigation, one of
the most elusive is undoubtedly to provide a Coulomb Gas (CG)
description for coupled Potts models. A putative CG construction would
allow one to compute the critical exponents of certain non-local
operators analytically, an issue that we only pursue numerically here.
Suffice it to say that our analytical result for the central charge at
the new self-dual critical point hints at a CG with three height
components, but only two of those are easy to identify geometrically.
This is not without reminding us of Ref.~\cite{Kondev98}.

The plan of the paper is as follows. In section \ref{sec:Potts} we
introduce the models to be studied, and summarize our main results. In
section~\ref{sec:critpoint} we derive the location and universality
class of a new self-dual critical point. The cornerstone of the
arguments is a level-rank duality in a related $SO(N)$ symmetric
integrable model. Numerical calculations are used to verify the
consistency of the proposed phase diagram, and to give results for the
dimensions of watermelon operators at the new critical point.  In
section~\ref{sec:critphase} we argue that adjacent to the new critical
point, the loop model possesses a critical phase, with
parameter-independent critical exponents. We provide evidence that the
phase transition into this phase is simultaneously first and second
order. In section~\ref{sec:dilute} we generalize the model to include
vacancies. We argue that dilution is an irrelevant perturbation of the
completely packed model. The phase diagram for the dilute model is
presented.  Finally, in section~\ref{sec:differentQ} we generalize our
results to the case where the two coupled Potts model have a different
number of states.

\section{The models and their phase diagrams}
\label{sec:Potts}

The original formulation of the $Q$-state Potts model places a spin
$\sigma_i$ taking integer values $1,2,\dots,Q$ on each site of some
lattice. In this paper we shall discuss exclusively the square
lattice, but we expect by universality that our results will apply
qualitatively and sometimes quantitatively to other lattices. The
Potts model has nearest-neighbor interactions invariant under the
permutation group $S_Q$, so that the interaction energy $J
\delta_{\sigma_i,\sigma_j}$ only depends on whether adjacent spins
$\sigma_i$ and $\sigma_j$ are the same or different.

The Potts model on the square lattice has a Kramers-Wannier (high- to
low-temperature) duality \cite{Potts52}. This is a non-local map of
spins which preserves the bulk free energy. The self-dual point is
critical \cite{Baxter73} when $Q\le 4$. In a field-theory description
of the region around these critical points, taking the temperature off
the self-dual value corresponds to perturbing by the energy operator
$\varepsilon$ (the continuum limit of $J \delta_{\sigma_i,\sigma_j}$).
The energy operator is odd under duality, $\varepsilon \to
-\varepsilon$, so that perturbing by this operator with one sign
describes the high-temperature phase, while the other sign gives the
low-temperature phase.

Although the original definition of the Potts model requires $Q$ to be
an integer, there exist generalizations to all $Q$ which are still
commonly known as the Potts model. We will treat two closely-related
but distinct models, which we refer to as the {\em loop} and the {\em
  height} models. The transfer matrices of the original Potts model
and the two generalizations are most conveniently defined in terms of
the Temperley-Lieb algebra; each model then corresponds to a
different representation. The algebra for a system of width $L$ has
$L$ generators $e_i$ acting at positions $i=1,2,\ldots,L$ as well as the
identity $I$, which obey the relations \cite{TL}
\begin{equation}
e_i^2=\sqrt{Q}e_i, \qquad e_i e_{i\pm 1}e_i = e_i, \qquad e_ie_j = e_j e_i
\hbox{ for }|i-j|>1 
\label{TLalg}
\end{equation}
The transfer matrix at the self-dual
isotropic point for width $L$ is then
\begin{equation}
 T=R_1R_3\dots R_{L-1}R_2R_4\dots R_L
\label{transfermatrix}
\end{equation}
where $R_i = I+e_i$ for the ferromagnetic self-dual point, and
$R=I-e_i$ for the (non-unitary) antiferromagnetic self-dual point. 
The partition function for an $M \times L$ lattice with periodic boundary
conditions in the ``time'' direction is then given by
\begin{equation}
Z=\hbox{tr } T^M.
\label{ZT}
\end{equation}
In this formulation, duality amounts to exchanging $e_i$ with $I$ on
each plaquette.

It is convenient to parameterize
\begin{equation}
 \sqrt{Q}=2\cos \left( \frac{\pi}{k+2} \right)
\label{paramk}
\end{equation}
defining the parameter $k$. When $k$ is an integer obeying $k\ge 2$
(so that $2\le Q \le 4$) representations of the algebra have special
properties. One important one discussed below is the existence of a
height representation, with positive and locally defined Boltzmann
weights. Having $k$ integer is also desirable for the quantum loop
models, because the existence of the Jones-Wenzl projector at these
values means that the number of ground states remains finite even when
the theory is defined on a torus \cite{Freedman01}.

\begin{figure}[ht]
\begin{center}
\includegraphics[width=.2\textwidth,angle=0]{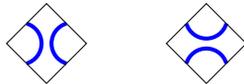}
\end{center}
\caption{Action of $I$ (left picture) and $e_i$ (right picture) on a
  pair of strands at positions $i$ and $i+1$. The transfer (time)
  direction is upwards. Each picture corresponds to a plaquette of the
  dual lattice.}
\label{fig:TL}
\end{figure}
\begin{figure}[ht]
\begin{center}
\includegraphics[width=14cm,angle=0]{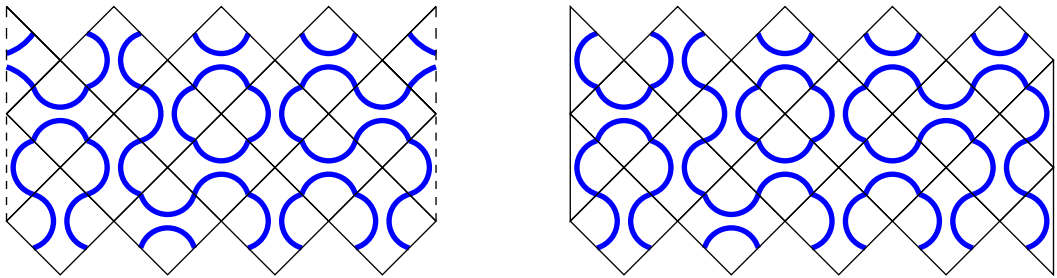}
\end{center}
\caption{Action of a term in $T^2$ on a zig-zag row of $L$ plaquettes.
  The left (resp.\ right) picture illustrates periodic (resp.\ free,
  or reflecting) boundary conditions in the space direction.}
\label{fig:transfer}
\end{figure}

In the {\em loop representation} of the Temperley-Lieb algebra, the
generators act on $L$ strands, as shown in Fig.~\ref{fig:TL}. The
result is a {\em completely packed} set of self-avoiding and
mutually-avoiding loops. Complete packing means that the loops cover
every link of the square lattice; the only degrees of freedom are the
two ways in which the loops avoid each other at each vertex of the
lattice.  The transfer matrix adds $L$ plaquettes to the dual lattice
as illustrated in Fig.~\ref{fig:transfer}, so it is acting on a
zig-zag row of $L$ plaquettes. Multiplying out the product in
(\ref{transfermatrix}) makes it obvious that each configuration in the
completely packed loop model appears with a coefficient which is just
$(\sqrt{Q})^{\#{\rm loops}}$ when $R_i = I+e_i$, and
$(-\sqrt{Q})^{\#{\rm loops}}$ when $R_i = I-e_i$.
Note in particular that the parameter $k$ in (\ref{paramk}) does not need
to be an integer for the loop representation to be defined.

A remarkable property of the loop representation is that the Boltzmann
weight of a configuration depends only on topological properties of
the loops. (In fact, for this reason the loop representation of the
Temperley-Lieb algebra underlies the Jones polynomial in knot theory.)
The local topological properties follow directly from the Temperley-Lieb
algebra (\ref{TLalg}). The second relation means that configurations
which are topologically equivalent receive the same weight, while he
first relation means that a closed loop not wrapping around a cycle
receives a weight $\sqrt{Q}$ relative to the configuration without the
loop.

To evaluate the full partition function in the loop representation, we
need to define the trace in (\ref{ZT}), which corresponds to imposing
periodic boundary conditions in the ``time'' direction. It is natural to
require that closed loops wrapping around the time direction also
receive a weight $\sqrt{Q}$. Thus the weight of a given loop
configuration ${\cal L}$ is found simply by counting the number
$n_{\cal L}$ of loops is the configuration:
$$Z=\sum_{\cal L} Q^{n_{\cal L}/2}$$
This defines the Markov trace of elements of the Temperley-Lieb algebra.
A loop is inherently a non-local object,
so even though the Boltzmann weights here are all positive for $Q\ge
0$, the corresponding field theory is typically not unitary.

When $k$ is an integer, the Temperley-Lieb algebra has a {\em height
  representation} where the resulting lattice model (often
known as RSOS, or restricted solid-on-solid, model) has positive
Boltzmann weights \cite{ABF,Huse,Pasquier87b,Pasquier87a}. The heights
$h_i=1,2,\ldots,k+1$ live on the dual lattice, so that each plaquette
illustrated in Fig.\ \ref{fig:transfer} 
has four heights at its corners.  Heights of neighboring sites
satisfy the RSOS constraint $|h_i-h_j| = 1$.  The generator $e_i$ now
acts locally as
\begin{equation}
 e_i |h_{i-1},h_i,h_{i+1} \rangle = \delta_{h_{i-1},h_{i+1}}
 \sum_{h'_i} \left( \frac{\sin\left( \frac{\pi h_i}{k+2} \right)
                         \sin\left( \frac{\pi h'_i}{k+2} \right)}
                        {\sin\left( \frac{\pi h_{i-1}}{k+2} \right)
                         \sin\left( \frac{\pi h_{i+1}}{k+2} \right)}
            \right)^{1/2} |h_{i-1},h'_i,h_{i+1} \rangle
\end{equation}
and the trace in (\ref{ZT}) is simply the usual matrix trace.

The connection between the loop and height models---valid when both
are defined, i.e., for $k$ integer---is based on exact partition
function identities \cite{Pasquier87b} that hold strictly speaking only
when the underlying lattice is {\em planar}. Imposing doubly periodic
boundary conditions leads to further subtleties, which
are however not important in the context of this work.

This paper is devoted to studying {\em two coupled self-dual Potts models},
in both their loop and height formulations.  We thus consider two
Potts models on the same lattice, described respectively by two
independent sets of Temperley-Lieb generators labeled by $e_i$ and
$f_i$. Here we require that $Q$ be the same for the two models; later
we will consider the more general case.  We require that the coupled
model remain self-dual under the combination of the duality
transformations in both models, and that it be invariant under
exchange of the two models. The transfer matrix remains of the form
(\ref{transfermatrix}), with $R_i$ for each plaquette given by
\begin{equation} 
 R_i = I\otimes I + e_i\otimes f_i + \lambda (I\otimes f_i + e_i\otimes I)
\label{coupled}
\end{equation}

\begin{figure}
\begin{center}
\includegraphics[width=10cm,angle=0]{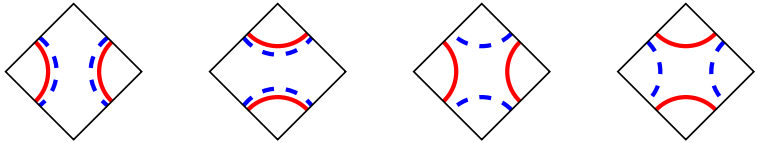}
\end{center}
\caption{The four configurations at each vertex in the coupled
  completely packed loop model.}
 \label{fig:4vertices}
\end{figure}

In the loop representation, this is very simple to represent; the four
possible configurations at each vertex are displayed in
Fig.~\ref{fig:4vertices}.  The two copies mean that on each link there
are two species of loops. Loops of each species cannot cross at
vertices, but the two species can cross. The partition function is a
sum over all configurations ${\cal L}$ and ${\cal M}$ of the two
species of loops:
\begin{equation}
Z= \sum_{{\cal L}} \sum_{{\cal M}} Q^{(n_{\cal L}+n_{\cal
M})/2}\lambda^{n_X}
\end{equation}
where $n_X$ is the number of times the rightmost two vertices in
Fig.~\ref{fig:4vertices} occur. This is a completely packed version of
the loop model discussed in \cite{Floops}; we will make contact with
the results discussed there when we discuss the dilute generalization
in section~\ref{sec:dilute}.

When the parameter $\lambda=1$ and $\lambda=-1$ , we have decoupled
ferromagnetic and antiferromagnetic models respectively, both
critical.  When $\lambda=0$, $R_i$ describes a single $Q^2$-state
Potts model, because the products $e_i \otimes f_i$ generate a
Temperley-Lieb algebra (\ref{TLalg}) with $Q$ replaced by $Q^2$. In
the loop representation, $\lambda=0$ amounts to requiring that the
loops in the two models be identical, so that each loop gets a weight
$(\sqrt{Q})^2=Q$. Taking the dual of one of the two models exchanges
$\lambda \mapsto 1/\lambda$, so we restrict our attention to $|\lambda|\le
1$.

In the height representation, the theory remains unitary for
$\lambda\ge -1/\sqrt{Q}$. This follows by explicitly examining the
Boltzmann weights; they all remain positive except for one type of
plaquette. Namely, a negative Boltzmann weight occurs for $\lambda<0$
for all plaquettes where the heights in one RSOS model differ in the
time direction, while the heights in the other RSOS model differ in
the space direction. Because of the periodic boundary conditions in
the time direction, there must be an even number of such
plaquettes. Thus one can flip the sign of these weights without
changing $Z$, giving a local model with all positive Boltzmann
weights. This remains true until $\lambda=-1/\sqrt{Q}$, where other weights
change sign. 

For $k=2$, perturbing $\lambda$ away from $1$ is marginal and
preserves the criticality. This is easy to see in the height
formulation, because $k=2$ corresponds to $Q=2$: (\ref{coupled}) then
describes two coupled Ising models. This is known as the Ashkin-Teller
model along its self-dual line, and the physics is well understood in
the height formulation \cite{Baxbook}. The model remains critical
until $\lambda=0$, where it is equivalent to the critical 4-state
Potts model. For $\lambda<0$, it is no longer critical.

At the decoupled points for $Q\le 4$, the models are critical and so
described by a conformal field theory.  In the height model for
$\lambda=1$, the conformal field theories describing these critical
models are known as the unitary minimal models, and have central
charge \cite{Huse}
\begin{equation}
 c = 2\left(1-\frac{6}{(k+1)(k+2)}\right) \qquad \mbox{for $\lambda=1$}
\label{cmin}
\end{equation}
The behavior of the decoupled height models at the antiferromagnetic
critical point $\lambda=-1$ is somewhat trickier to describe. The
negative Boltzmann weights make the theory non-unitary. However, if
one instead uses a Hamiltonian formulation of a 1+1 dimensional
quantum theory, the Hamiltonian corresponding to $\lambda=-1$ is
simply $H=\sum_{i=1}^L e_{i}$, which is hermitian. The unitary
conformal field theory describing the continuum limit of this
Hamiltonian is known as the $Z_k$ parafermion model, whose central
charge is
\begin{equation}
 c_{\rm height} = 2 \left(\frac{2(k-1)}{k+2} \right)
\qquad \mbox{for $\lambda=-1$.}
 \label{cheight-1}
\end{equation}

In the loop model, (\ref{cmin}) remains true for $\lambda=1$. However,
since $k$ is now a continuous parameter, the cases $\lambda=1$ and
$\lambda=-1$ are now linked by analytic continuation, $k \mapsto
-\frac{k}{k+1}$. Note that under this transformation $\sqrt{Q} \mapsto
-\sqrt{Q}$, which is tantamount to $R_i = I+e_i \mapsto I-e_i$ as
required.  In particular, for $\lambda=-1$ the correct central charge
is then obtained by analytically continuing (\ref{cmin}) and reads
\begin{equation}
 c_{\rm loop} = 2 \left(
  1 - \frac{6(k+1)^2}{k+2} \right) \qquad \mbox{for $\lambda=-1$.}
 \label{cloop-1}
\end{equation}
When $k$ is integer, the corresponding critical theory completely
disappears from $Z$ due to massive cancellations \cite{Saleur91} of
eigenvalues in the Markov trace (\ref{ZT}), and what remains is
tantamount to the height model.

Taking $\lambda$ away from $1$ is a perturbation odd under each the
individual dualities of the Potts models, and even under the
combination. The simplest operator with these properties is given by
the product of the energy operators $\varepsilon_1 \varepsilon_2$ of
the two models. This operator has dimension $(k+4)/(k+1)$, so it is
relevant for $k>2$; it is the only relevant one with the appropriate
symmetries. Thus the question becomes whether this perturbation causes
a flow to a non-trivial critical point. Compelling evidence suggests
that there is no such critical point, although no definitive argument
rules it out, Namely, this perturbation is integrable in the field
theory \cite{Vaysburd}, and the only consistent $S$ matrices
describing this perturbed theory are massive, indicating there is no
flow to a critical point. \footnote{We note, however, that the $S$
matrices given in \cite{Vaysburd} for $k>2$ are wrong---they are the
level-rank duals of the correct ones, in the technical sense described
below.}  Moreover, this is consistent with the fact that at
$\lambda=0$, the model for $k>2$ is non-critical, because the
$Q^2$-state Potts model at its self-dual point is critical only for $Q
\le 2$. Thus it is natural to guess that for $k>2$, the flow goes from
$\lambda=1$ to $\lambda=0$, the latter being a stable non-critical
fixed point of the renormalization group. Our numerical work supports
this picture.

The interesting new physics occurs for $\lambda$ negative, where we
find a new critical point. We present this calculation in section
\ref{sec:critpoint}, where we derive the exact location of the
critical point and determine which conformal field theory describes
its continuum limit.

\begin{figure}
\begin{center}
\includegraphics[width=.7\textwidth,angle=0]{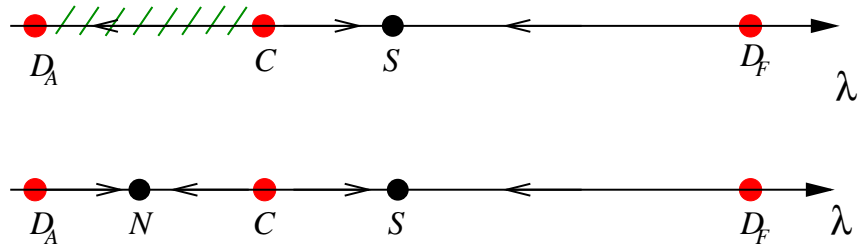}
\end{center}
\caption{Phase diagram for the loop model (top) and
  the height model (bottom). The arrows indicate
  renormalization group flows. The cross-hatching between the new
  critical point $C$ and the decoupled antiferromagnetic critical
  point $D_A$ represents the critical phase.}
 \label{fig:phase_dense}
\end{figure}

The arguments given above---in conjunction with numerical calculations
presented in section~\ref{sec:numdense} below---lead to the phase
diagrams shown in Fig.~\ref{fig:phase_dense}. The point $S$ is the
strongly coupled point $\lambda=0$, which is non-critical for
$k>2$. The point $C$ is the new critical point at $\lambda_c$, to be
discussed in detail in section~\ref{sec:critpoint} below. The points
$D_F$ and $D_A$ are the decoupled ferromagnetic and antiferromagnetic
critical points at $\lambda=1$ and $\lambda=-1$. The characteristics
of $D_A$ differ between the height and the loop models. Finally, the
point $N$ is a non-critical point which occurs only in the height
model, and which is required for the consistency of its RG flow
diagram.

Note in particular that the phase $\lambda \in [-1,\lambda_c)$ is
non-critical in the height model, but critical in the loop model. The
latter is an example of a ``Berker-Kadanoff'' phase \cite{Saleur91}.

The point $\lambda=-1/\sqrt{Q}$, the lowest value of $\lambda$ where
the height representation is unitary, is special. It is quite
important for quantum loop models, and it lies in the critical phase
for $k<6$, and is the new critical point for $k=6$. When
$\lambda=-1/\sqrt{Q}$, the weight of each configuration can explicitly
be given in terms of a chromatic polynomial squared
\cite{Fendley08}. This can be seen for the square lattice by noting
that
\begin{eqnarray*}
R&=& (1-e/d)(1-f/d) + (d^2-1)(e/d)(f/d)\\
&=& (e-1/d)(f-1/d) + (d^2-1)(1/d^2)
\end{eqnarray*}
where $d=\sqrt{Q}$. The projectors $1-e/d$ and $e/d$ are orthogonal.
We divide the square lattice on which the loops live into the usual
two sublattices (even and odd).  The {\em low}-temperature expansion
of a single Potts model is found from writing $R_{\rm Potts}=(1-e/d)
+(d-1)e/d$ on sublattice 1, and $R_{\rm Potts}=(e-1/d)+(d-1)/d$ on
sublattice 2, and then associating a domain wall with each $1-e/d$
term included on sublattice 1, and $e-1/d$ on sublattice 2. It is easy
to check that these walls cannot end, although there can be three or
four walls touching each site.  If we sum over the spins for any fixed
domain-wall configuration, the weight per graph is proportional to the
chromatic polynomial of the dual lattice. This is easiest to see by
simply proving it for all $Q$ integer in the original Potts
representation; by definition of the chromatic polynomial it must
apply for all $Q$.  Thus coupling Potts models at $\lambda=-1/d$
corresponds to requiring that the low-temperature expansion of the two
be the same. (This is proved for any lattice with four links touching
each site in \cite{Fendley08}.) The behavior of the low-temperature
expansion at $\lambda=-1/d$ is therefore analogous to that of the
completely packed (Fortuin-Kasteleyn \cite{FK}) loop expansion at
$\lambda=0$. 

\section{The new self-dual critical point}
\label{sec:critpoint}

In this section we show that at a particular value
$\lambda=\lambda_c$, the coupled Potts models have a critical
point. In accord with the discussion of the previous section, this
value must be negative, and we show here that
\begin{equation}
\lambda_c = -\sqrt{2}\sin\left(\frac{\pi(k-2)}{4(k+2)}\right) \,.
\label{lamc}
\end{equation}
The conformal field theory describing this critical theory is the
coset model \cite{GKO} 
\begin{equation}
\frac{SU(2)_k \times SU(2)_k}{SU(2)_{2k}}
\label{coset}
\end{equation}
with central charge
\begin{equation}
c=\frac{3k^2}{(k+1)(k+2)} \,.
\label{ck}
\end{equation}

\subsection{Level-rank duality and the BMW algebra}

At the value $\lambda=\lambda_c$, the coupled theory is integrable.
We show this by using level-rank duality to write the transfer matrix
here in terms of another algebra, the $SO(N)$ Birman-Murakami-Wenzl
(BMW) algebra \cite{BMW}.  Lattice models involving this algebra were
introduced long ago \cite{JMO}, and it is possible to show that
integrability follows directly from algebraic properties, and so
applies to any representation \cite{Akutsu}. In this more general
setting, the Andrews-Baxter-Forrester height models \cite{ABF}
discussed are associated with $SU(2)$ and its quantum-group algebra
deformation $U_q \big( SU(2) \big)$.

The BMW algebra was introduced to study representations of the
braid group and find new knot and link invariants. 
Braid generators in general
satisfy $B_iB_j = B_jB_i$ for $|i-j|>1$, and 
\begin{equation}
B_i B_{i+1} B_i = B_{i+1} B_i B_{i+1}\ .
\label{reid1}
\end{equation}
This is known as the ``third Reidemeister move'' in knot theory.  The
braid generators can be inverted: $B_i B_i^{-1}=I$ (the second
Reidemeister move).  A useful graphical representation of the braid
group is given by each $B_i$ and $B_i^{-1}$ as acting on two strands
like the Temperley-Lieb generators, but where $B$ and $B^{-1}$
correspond to overcrossings and undercrossings respectively. 

The $SO(N)$ BMW algebra is generated by the $B_i$ and $B_i^{-1}$,
where $i=1,2,\ldots,L$. This algebra involves a parameter $q$ as well
as $N$; for the models we label $SO(N)_k$, $q=e^{i\pi/(N+k-2)}$ is a
root of unity.  The braid generators of the BMW algebra of course
satisfy the second and third Reidemeister moves, but the algebra
has more structure. The Temperley-Lieb algebra is a subalgebra, with
$E_i$ defined as
\begin{equation}
E_i = I + \frac{B_i - B_i^{-1}}{q-q^{-1}}
\label{EIB}
\end{equation}
where $q$ is a parameter. Here the $E_i$ obey
$$(E_i)^2=DE_i,\qquad E_iE_{i+1}E_i =E_i,\qquad E_iE_j=E_jE_i \hbox{
  for } |i-j|>1$$
where
$$D= 1 + \frac{q^{N-1}-q^{1-N}}{q-q^{-1}}\ .$$
The full set of BMW relations can be found in \cite{Akutsu}. 
One useful
one is the first Reidemeister move
$$B_i E_i = E_i B_i = q^{N-1} E_i\ .$$
For later use, it is
convenient to define
\begin{equation}
X_i = q^{-1}I +q E_i - B_i
\label{Xdef}
\end{equation} 
This generator $X_i$ is particularly interesting in the $N=3$ case,
where the BMW algebra is related to the chromatic algebra
\cite{fendleyread,FKrush}. 

The coupled Potts model transfer matrix (\ref{coupled})
can be written in terms of the $SO(4)_k$ BMW
algebra. This BMW algebra is special
because the algebra $SO(4)$ is the direct product of two $SU(2)$
algebras. One should think of the strands in $SO(N)_k$ as behaving in
the vector representation of the quantum-group algebra
$U_q(SO(N))$. As with ordinary Lie algebras, $U_q(SO(4))$ 
decomposes into copies of $U_q(SU(2))$, so the lines in $SO(4)_k$  behave as
two independent spin-1/2 lines in $SU(2)_k$. Spin-1/2 lines obeying
the Temperley-Lieb algebra satisfy the braid group relation
$b_ib_{i+1}b_i=b_{i+1}b_i b_{i+1}$ and $b_i b_i^{-1}=I$ if we
define the $b_i$ via the ``skein relation''
$$b_i = q^{-1/2}-q^{1/2}e_i,\qquad
b_i^{-1} = q^{1/2}-q^{-1/2}e_i,\qquad
$$ 
where the $e_i$ obey (\ref{TLalg}) for $\sqrt{Q}=q+q^{-1}$. (This
is the braid-group representation underlying the Jones
polynomial.) So less abstractly, given two independent
representations labeled $e_i$ and $f_i$ of the TL algebra, we can
construct one of the $SO(4)_k$ BMW algebra by defining
\begin{eqnarray}
E^{(4)}_i &=& e_i\otimes f_i
\label{BMWE4}\\
B^{(4)}_i &=& (q^{-1/2}I-q^{1/2}e_i)\otimes
(q^{-1/2}I-q^{1/2}f_i)=q^{-1}I + qE^{(4)}_i - X^{(4)}_i
\label{BMWB4}
\end{eqnarray}
where given the definition of $X_i$ in (\ref{Xdef}), we have
$$X^{(4)}_i= e_i\otimes I  +I\otimes f_i$$
for this representation of $SO(4)_k$.
Therefore the transfer matrix (\ref{coupled}) for each plaquette can be
written simply as
\begin{equation}
R_i = I + E^{(4)}_i +\lambda X^{(4)}_i\ .
\label{RSO4}
\end{equation}
Because of the relation (\ref{EIB}), we need use only three of the
generators for each plaquette.

Lattice models whose transfer matrices are built on the $SO(N)_k$ BMW
algebra are integrable only for special values of the parameters.
Height models associated with all non-exceptional Lie algebras were
introduced in \cite{JMO}.  The transfer matrices for the critical
points of these lattice models can be written in terms of BMW
algebras; for the $SO(N)_k$ case they are \cite{Akutsu}
\begin{equation}
R^{(N)}_i(u)=[\gamma-u][1-u]I + [u][1-\gamma+u]E_i + [\gamma-u][u] X_i
\label{Ru}
\end{equation}
where $\gamma=(N-2)/2$ and we have introduced the shorthand
$$[x]= \sin\left(\frac{\pi x}{N+k-2}\right)= \frac{1}{2}(q^x-q^{-x})\ .$$
The parameter $u$ is simply a lattice anisotropy which does not affect
any of the universal properties. Under 90 degree rotations, $I$ and
$E$ are interchanged, while $X$ remains invariant. Thus isotropic
Boltzmann weights are found by setting $u=\gamma/2$, giving
\begin{equation}
\frac{R^{(N)}_i(\gamma/2)}{[\gamma/2][1-\gamma/2]} = I + E_i +
\frac{[\gamma/2]}{[1-\gamma/2]} X_i
\label{Riso}
\end{equation}
{}From the BMW algebra it follows that 
this $R$ matrix satisfies the
Yang-Baxter equation 
$$R_i(u)R_{i+1}(u+u')R_i(u')=R_{i+1}(u')R_i(u+u')R_{i+1}(u) \,,$$
so that the model is integrable. Note that the braid generators are
recovered in the $u\to\pm\infty$ limits:
$$\lim_{u\to \pm\infty} \frac{R^{(N)}_i(u)}{q^{\pm (u-\gamma)}}= q^{\mp 1}I +
  q^{\pm 1} E_i -X_i = B_i^{\pm 1}\ .$$ 

It is straightforward to establish that lattice models with the
transfer matrix given by (\ref{Ru}) and (\ref{transfermatrix}) are
critical. This is a generic feature of integrable models with
trigonometric Boltzmann weights. Here one can see this by utilizing a
deformation of these models which preserves the integrability
\cite{JMO}. This enables one for example to compute the free energy
by using the inversion-relation method \cite{Baxbook}. From this one
sees that as the deformation parameter is tuned back to the
trigonometric weights, the free energy behaves non-analytically, as
it does at a critical point. The central charge of the conformal
field theory describing this critical point can be computed after making some
(very believable) technical assumptions; for $u>0$ it is
\cite{KNS}
$$c=\frac{N}{2} \left( 1 - \frac{N^2-3N+2}{(N+k-2)(N+k-3)} \right)$$
which is the central charge of the
\begin{equation}
\frac{SO(N)_{k-1} \times SO(N)_1}{SO(N)_{k}}
\label{cosetSON}
\end{equation}
coset models.

The new critical point for coupled Potts models does {\em not}
correspond to the $N=4$ case of (\ref{Riso}). Rather, this yields the
decoupled ferromagnetic critical point. Namely, when $N=4$, $\gamma=1$,
and the decomposition of the $SO(4)$ BMW algebra in terms of two
Temperley-Lieb algebras in (\ref{BMWE4},\ref{BMWB4})
means 
$$R^{(4)}_i(u)=([1-u]I + [u]e_i)\otimes ([1-u]I + [u]f_i)$$
Thus the models decouple here: at the isotropic point $u>0$ this is
simply the $\lambda=1$ ferromagnetic critical point. Indeed, because
as conformal field theories $SO(4)_k=SU(2)_k\times SU(2)_k$, the coset
in (\ref{cosetSON}) simply splits into two copies of
$$\frac{SU(2)_{k-1} \times SU(2)_1}{SU(2)_k}$$
with central charge (\ref{cmin}). 


To find the new critical point at $\lambda=\lambda_c$, we need to exploit
the {\em level-rank duality} of the BMW algebra \cite{FKrush}. It is simple to
check that if the $B_i$ satisfy this $SO(N)_k$ algebra, then the
$-B_i^{-1}$ satisfy the $SO(k)_N$ algebra. This follows because $q$ is
invariant under this exchange, and
$$q^{N-1}q^{k-1} = -1\ $$
so that $D$ remains unchanged. 
Thus from any representation of $SO(N)_k$ with generators $\{B_i,E_i\}$, 
we can construct one of
$SO(k)_N$ with generators $\{\widetilde{B}_i,\widetilde{E}_i\}$ by
$$\widetilde{B}_i = - B_i^{-1}, \qquad \widetilde{E}_i=E_i.$$
We then have
\begin{eqnarray*}
\widetilde{X}_i &=& q^{-1}I + q E_i + B_i^{-1}\\
&=& q^{-1}I + qE_i + (q-q^{-1})(I-E_i)+B_i\\
&=& (q+q^{-1})(I+E_i) -X_i
\end{eqnarray*}

The transfer matrix of the coupled Potts models defined in
(\ref{coupled}) has been expressed in terms of the $SO(4)_k$ BMW
algebra via (\ref{RSO4}). The level-rank duality therefore implies
that it can also be written in terms of the generators
$\{\widetilde{E}_i,\widetilde{X}_i\}$ of the $SO(k)_4$ BMW algebra:
\begin{equation}
R_i = (1+\lambda(q+q^{-1}))(I+\widetilde{E}_i) -\lambda \widetilde{X}_i 
\label{RXtilde}
\end{equation}
We can now find the new critical point, because the
$SO(k)_4$ models are critical
when the Boltzmann weights obey (\ref{Riso}).  It
follows that the coupled Potts models are critical at
$\lambda=\lambda_c$, where $R_i\propto R^{(k)}_i$. This holds if 
$$-\frac{\lambda_c}{1+\lambda_c(q+q^{-1})} =
\frac{[\gamma/2]}{[1-\gamma/2]}.$$ 
Plugging in the appropriate values
of $[\gamma/2]$ and $[1-\gamma/2]$ for $SO(k)_4$ and solving for
$\lambda_c$ gives (\ref{lamc}).

We expect the central charge to be independent of the representation
of the BMW algebra used, as long as appropriate boundary conditions
are chosen. This critical point at $\lambda=\lambda_c$
therefore should have the central charge of the coset conformal field
theory
$$\frac{SO(k)_3\times SO(k)_1}{SO(k)_4} \,,$$ which is that given in
(\ref{ck}). A related type of level-rank duality arises in coset
conformal field theories \cite{Alt}. This lets us map this coset model
on to others:
$$\frac{SO(k)_3\times SO(k)_1}{SO(k)_4} \approx
\frac{SO(4)_k}{SO(3)_k} = 
\frac{SU(2)_k\times SU(2)_k}{SU(2)_{2k}}\ .$$ 
The latter coset is the one we gave above in (\ref{coset}); note
that the coset description involving $SO(4)$ is {\em not} the $SO(4)$
case of (\ref{cosetSON}). Thus exploiting the integrability of this
lattice model lets us not only determine the location of the critical
point exactly, but it lets us determine the correct coset conformal
field theory.

Note that when $N=k$, the level-rank duality does not change the
model, only the couplings. The points $\lambda=1$ and
$\lambda=\lambda_c$ are not self-dual under level-rank duality, but rather
are mapped to each other. Thus for the coupled Potts models with
$N=k=4$ ($Q=3$), the new critical point at $\lambda=\lambda_c$ is
equivalent to the usual $\lambda=1$ pair of decoupled models. Indeed,
the central charge (\ref{ck}) here at $\lambda=\lambda_c$ is $8/5$,
that of two decoupled 3-state Potts models at their ferromagnetic
critical points. As pointed out in section 5.2.2 of \cite{DJLP}, the coupled
$Q=3$ models in fact have an exact symmetry $J \mapsto -J$,
where $J$ is the original Potts spin coupling. In the present
notation we have $\lambda = ({\rm e}^J-1) / \sqrt{Q}$, and the
symmetry reads $\lambda \mapsto - \frac{\lambda}{1+\lambda
\sqrt{Q}}$. For $Q=3$ this gives indeed $1 \mapsto -
\frac{1}{1+\sqrt{3}} = \lambda_c$.

\subsection{Numerical calculations}
\label{sec:numdense}

The coupled model can be studied numerically using techniques for
exactly diagonalizing the transfer matrix $T$, either in the loop or the
height representation. We focus here on the loop model.

For reasons of numerical efficiency, it turns out to be most convenient
to chose the transfer direction to be axial with respect to the square
lattice on which the loops live (as opposed to the diagonal transfer
direction depicted in Figs.~\ref{fig:transfer}--\ref{fig:4vertices}).
Periodic boundary conditions are imposed in the space direction.
To avoid parity effects in the antiferromagnetic region, the number of
strands $L$ is chosen to be even. We can then handle widths up to
$L=18$.

\begin{figure}
\begin{center}
\includegraphics[width=10cm,angle=270]{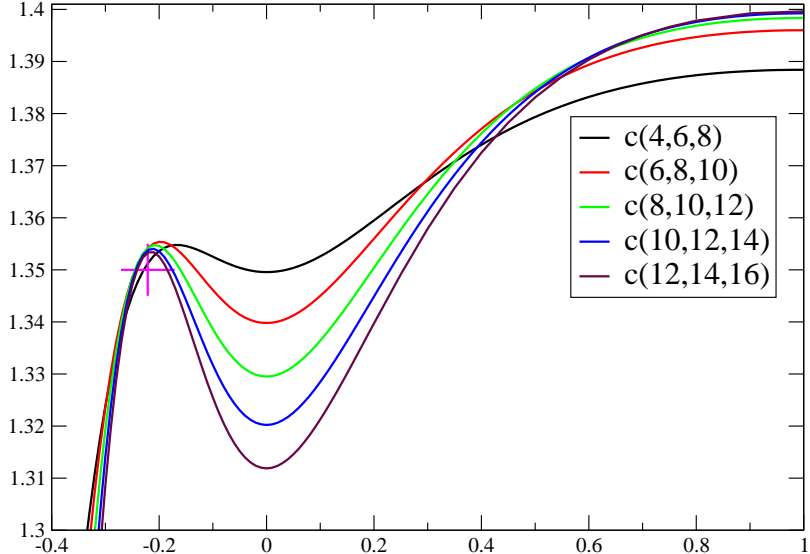}
\end{center}
\caption{Central charge of the loop model with $k=3$, as a function of
  $\lambda$. The pink cross indicates the expected infinite-size
  behavior at the new self-dual critical point.}
 \label{fig:centralB5}
\end{figure}

The central charge can be extracted from finite-size scaling of the
leading eigenvalue of $T$ in the standard way. The resulting
three-point fits $c(L-4,L-2,L)$ are shown in Fig.~\ref{fig:centralB5}
as functions of $\lambda$. The figure shows the case $k=3$, but larger
integer values of $k$ lead to similar results. Eq.~(\ref{cmin}) then predicts
$c=\frac{7}{5}=1.4$ at the decoupled ferromagnetic point $\lambda=1$,
and Eq.~(\ref{ck}) gives $c=\frac{27}{20}=1.35$ at the new self-dual
critical point, which according to (\ref{lamc}) is situated at
$\lambda_c = -\sqrt{2} \sin(\frac{\pi}{20}) \simeq -0.2212$.  Finally,
for the strongly coupled point $\lambda=0$ we have a single $Q^2 =
\left( 2 \cos(\frac{\pi}{5}) \right)^4 \simeq 6.8541$ state Potts
model, which is non-critical but with a rather large correlation
length.

All of this is nicely brought out by Fig.~\ref{fig:centralB5}.
The fact that the peaks in $c$ at the critical points $\lambda=\lambda_c$
and $\lambda=1$ narrow upon increasing $L$, further indicates that both
points are RG repulsive.

To summarize this far, our numerical results not only confirm the
existence of the critical point at $\lambda=\lambda_c$, but also
provide compelling evidence that the central charge (\ref{ck}) is the
correct one.

We now turn to the operator content of the critical theory.  One of
the most interesting and geometrically appealing operators that can be
defined in a loop model is the so-called $\ell$-leg watermelon
operator ${\cal O}_\ell$ \cite{Nienhuis}. To define this operator
properly, we first imagine assigning an arbitrary orientation
(clockwise or counterclockwise) to each loop in the model; the
original loop model is then recovered by independently summing over the
two orientations of each loop. The oriented loop model satisfies a
zero-divergence constraint: the net arrow flow out of any simply
connected region vanishes. The operator ${\cal O}_\ell$ breaks this
constraint, and corresponds to a point (or a small neighborhood of a
point) that acts as a source of $\ell$ oriented loop segments. The
critical exponents $x_\ell$ corresponding to the asymptotic decay of
the two-point correlation functions
\begin{equation}
 \langle {\cal O}_\ell({\bf x}_1) {\cal O}_{-\ell}({\bf x}_2) \rangle \sim
 |{\bf x}_1 - {\bf x}_2|^{-2 x_\ell} \qquad
 \mbox{for $|{\bf x}_1-{\bf x}_2| \ll 1$}
\end{equation}
can be computed analytically when the loop model admits a Coulomb gas
construction.  This has been accomplished for the loop model
corresponding to a single Potts model \cite{Nienhuis} as well as for
more complicated loop models \cite{Kondev98}.

The context of ${\cal N}$ coupled loop models allows for generalizing
these operators to $(\ell_1,\ell_2,\ldots,\ell_{\cal N})$-leg
watermelon operators ${\cal O}_{\ell_1,\ell_2,\ldots,\ell_{\cal
    N}}({\bf x})$, where the $p$th model supports an $\ell_p$-leg
defect at the point ${\bf x}$. The interpretation of this
generalization in the replica setup was given in \cite{DJLP}.  The
corresponding critical exponents can be measured numerically by
defining a modified transfer matrix in which the $p$th model is
constrained to having precisely $|\ell_p|$ loops which wind around the
time direction.

\begin{table}
\begin{center}
\begin{tabular}{l|lll}
 Exponent & $k=3$    & $k=4$    & $k=5$ \\ \hline
 $x_{2,0}$ & 0.998(1) & 0.999(1) & 0.999(1) \\
 $x_{2,2}$ & 0.200(2) & 0.130(2) & 0.095(2) \\
 $x_{4,0}$ & 1.560(2) & 2.5  (2) & ---      \\
 $x_{4,2}$ & 0.774(1) & 0.510(5) & 0.392(4) \\
 $x_{4,4}$ & 0.926(4) & 0.677(2) & 0.555(5) \\
 $x_{6,0}$ & 2.40 (3) & 2.47 (1) & 3.0  (1) \\
 $x_{6,2}$ & 1.70 (5) & 1.15 (5) & 0.84 (5) \\
 $x_{6,4}$ & 1.85 (5) & 1.28 (5) & 1.00 (5) \\
 $x_{6,6}$ & 2.08 (7) & 1.54 (5) & 1.26 (6) \\
\end{tabular}
\end{center}
\caption{Watermelon exponents $x_{\ell_1,\ell_2}$ at the critical point
 $\lambda_c$ for $k=3,4,5$. The error bar on the last quoted digit is
 given in parentheses. One entry is missing due to convergence problems
 in the diagonalization method being used.}
\label{tab:watermelons}
\end{table}

Returning to the case ${\cal N}=2$, we have determined numerically
the exponents $x_{\ell_1,\ell_2}$ (with $\ell_p=0,2,4,6$ for $p=1,2$)
at the critical point (\ref{lamc}). This study involved strips of
width up to $L=16$ strands. Results for parameter values $k=3,4,5$
are shown in Table~\ref{tab:watermelons}. These results give very
strong support for the following conjectures
\begin{equation}
 x_{2,0} = 1 \,, \qquad
 x_{2,2} = \frac{4}{(k+1)(k+2)}
\end{equation}
and moderately strong support for the additional conjecture
\begin{equation}
 x_{4,2} = \frac{16}{(k+1)(k+2)} \,.
\end{equation}
The denominator $(k+1)(k+2)$ of these expressions is in accord with heuristic
arguments; operators arising in the unitary coset conformal field
theories (\ref{coset}) describing these critical points indeed have
dimensions with this denominator. 

\section{The critical phase}
\label{sec:critphase}

The phase diagram for $\lambda<\lambda_c$ is different for the height
and loop models. To understand this, let us first study the region
around $\lambda=-1$, perturbing away from the decoupled
antiferromagnetic self-dual critical point. The situation is much
subtler than at $\lambda=1$, because the models for $\lambda<-1/d$ are
not unitary. We argue here that the perturbation around $\lambda=-1$
is relevant for the height model, but irrelevant for the loop
model.

Recall that the operator $\varepsilon_1 \varepsilon_2$ responsible for
the flow away from $\lambda=1$ has dimension $(k+4)/(k+1)$, and so is
relevant for $k>2$. In the loop model, the points $\lambda=1$ and
$\lambda=-1$ are linked by the analytic continuation $k \mapsto
-\frac{k}{k+1}$, and so $\varepsilon_1 \varepsilon_2$ has dimension
$3k+4$ and is irrelevant for any $k \ge 0$.  The
decoupled point $\lambda=-1$ should therefore act as an RG attractor
for a whole range of parameter values $\lambda \in [-1,\lambda_0)$,
i.e., we have a critical phase with constant values of the critical
exponents.

\begin{figure}[ht]
\begin{center}
\includegraphics[width=9.4cm,angle=270]{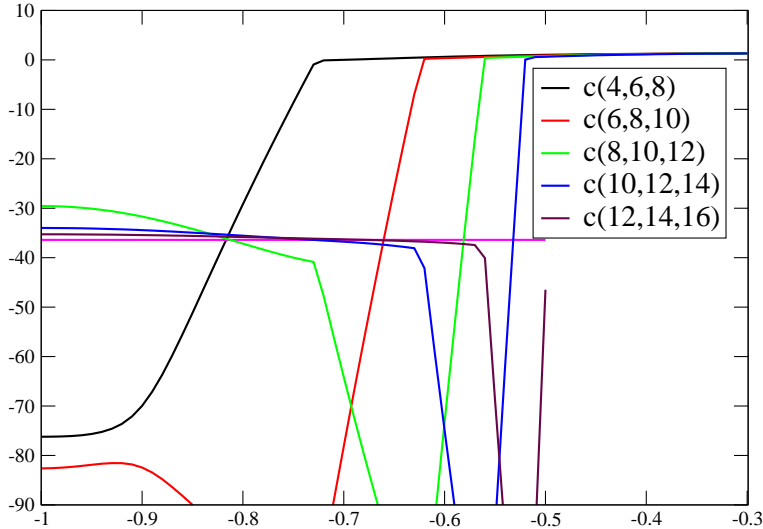}
\end{center}
\caption{Central charge of the loop model with $k=3$, as a function of
  $\lambda$. The pink horizontal line indicates the expected
  infinite-size behavior throughout the critical phase.}
 \label{fig:plateauB5}
\end{figure}

The existence of a critical phase in the loop model is verified
numerically in Fig.~\ref{fig:plateauB5} for the case $k=3$. The
central charge then reads, according to (\ref{cloop-1}),
$c=-\frac{182}{5} = -36.4$, in fine agreement with the numerics
despite of the large corrections-to-scaling effects.

\begin{figure}[ht]
\begin{center}
\includegraphics[width=9.4cm,angle=270]{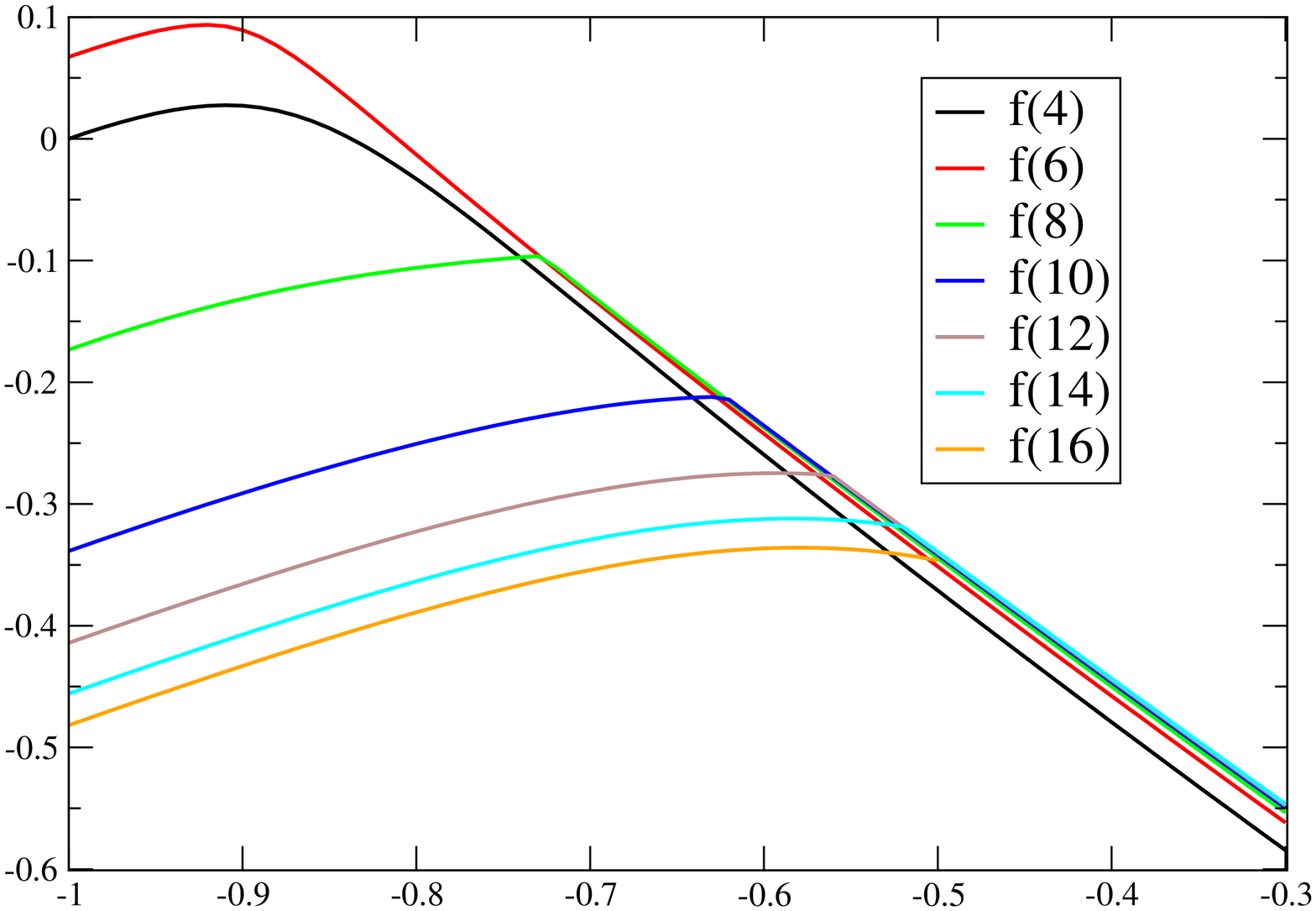}
\end{center}
\caption{Free energies $f(L)$ of the loop model with $k=3$, as functions of
  $\lambda$. Level crossings are visible for strip widths $L \ge 8$.}
 \label{fig:freeB5}
\end{figure}

We next examine more closely the extent of the critical phase, and the
nature of the transition at $\lambda_0$.  Fig.~\ref{fig:freeB5}
shows (still for $k=3$) the free energy per unit area $f(L) =
-\frac{1}{L} \log \lambda_{\rm max}$ for various strip widths $L$,
where $\lambda_{\rm max}$ is the dominant eigenvalue of the
corresponding transfer matrix. For $L \ge 8$ there is a jump
singularity in the derivative $f'(\lambda)$ at some
$\lambda=\lambda_0(L)$, signaling a first-order phase transition. It
is this singularity which marks the termination of the critical phase.

The existence of this first-order phase transition can be argued
analytically as well \cite{Salas}. Indeed, at $\lambda=-1$ the $k
\mapsto \tilde{k} \equiv -\frac{k}{k+1}$ transformation implies that
the dominant eigenvalue of the transfer matrix corresponds to a sector
of non-trivial topological charge. Specifically, the smallest (i.e.,
most {\em negative}) critical exponent corresponds, for $0 < \tilde{k}
< 2$, to a single Potts cluster wrapping around {\em both} the space
and time directions of the lattice (and thus there are {\em zero}
wrapping cluster boundaries), while for $\tilde{k} > 2$ it corresponds
to two Potts clusters (hence {\em four} cluster boundaries) wrapping
around the time direction. Needless to say, in the probabilistic
regime ($\lambda > 0$) the dominant eigenvalue corresponds to the
identity operator, and so no cluster wraps around the time direction
(for finite $L$, and in the limit $M \to \infty$). We conclude that
for some $\lambda_0(L) \in (-1,0)$ there must be a level crossing in
the transfer matrix spectrum, corresponding to a first-order
transition.

For $k=3$ we have numerically determined
$$\lambda_0(L)  = -0.72793, -0.62225, -0.56084, -0.51985, -0.49002
\qquad \mbox{for $L=8,10,12,14,16$.}$$
In the $L\to\infty$ limit this extrapolates to $\lambda = -0.25 \pm
0.05$, in conspicuous agreement with $\lambda_c \simeq -0.2212$ from
(\ref{lamc}). Similar agreements are found for other values of $k$.

We are therefore led to conjecture that $\lambda_0 = \lambda_c$ for
any $k$.  This means that at the critical point $\lambda_c$ the loop
model stands at a {\em continuous first-order phase transition}
\cite{Sokal}, while the height model stands at a conventional
second-order transition. The two transitions are in the same
universality class. This scenario is closely analogous to the one
found for the antiferromagnetic transition in a single Potts model
\cite{SaleurAF,IkhlefAF}.

Turning now to the height model, a perturbation away from $\lambda=-1$
in the $Z_k$ parafermion theory (\ref{cheight-1}). As with the
perturbation around $\lambda=-1$, the perturbing operator here must be
odd under the individual dualities of the models, and even under the
combination. In the unitary Hamiltonian formulation, this corresponds
to the product of energy operators in the parafermion theories, which
is relevant with dimension $4/(k+2)$.  Accordingly, we expect a
different scenario in which the critical phase does not
exist. Numerical transfer matrix diagonalization reveals that the free
energies $f(L)$ are smooth functions all the way down to $\lambda=-1$.
For $\lambda > \lambda_0(L)$ they are identically equal to those of
the loop model. Exactly at $\lambda=-1$ there is a jump singularity in
the derivative $f'(\lambda)$, as is consistent with the overall
$\lambda \mapsto 1/\lambda$ symmetry of the model. [At $\lambda=1$ the
Perron-Frobenius theorem prevents such a singularity from occurring.]
In between $\lambda=-1$ and $\lambda=\lambda_c$ the behavior is
non-critical, the effective central charge being zero.

\section{Dilute loops}
\label{sec:dilute}

The loop models we have studied thus far are completely packed: every
site of the lattice is covered by strands. It is natural to relax this
constraint, and consider ``dilute'' loops. We show in this section
that the critical point and phase still persist for fairly large
amounts of dilution. 

\begin{figure}[ht]
\begin{center}
\includegraphics[width=12cm,angle=0]{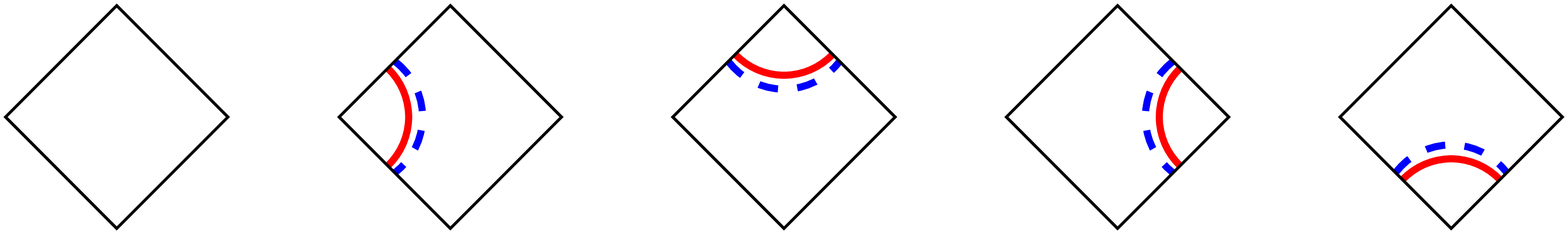}
\end{center}
\caption{The five additional configurations at each vertex in the
  dilute loop model.}
 \label{fig:5vertices}
\end{figure}

The dilute model we discuss here requires that loops of the two
species be on the same links: they are ``glued'' together. This does
not reduce to a single loop model, because when two loops meet at the
same vertex, the four possibilities in Fig.~\ref{fig:4vertices} are
still possible. For reasons we will discuss below, the physics does
not seem to be changed by whether or not one allows vertices with the
pair of loops going straight through. Thus for simplicity, we still
require that loops turn by 90 degrees at each vertex, so that there
are nine allowed configurations at each vertex. The five new
possibilities are displayed in Fig.~\ref{fig:5vertices}: in addition
to the empty vertex, there are four vertices where two links touching
each vertex are empty.

Let the weight of the vertices in Fig.~\ref{fig:5vertices} (from left
to right) be
\begin{equation}
 \omega_1=1 \quad \mbox{and} \quad \omega_2=\omega_3=\omega_4=\omega_5=K
\end{equation}
and those of the vertices in Fig.~\ref{fig:4vertices} (from left to right) be
\begin{equation}
 \omega_6=\omega_7=K^2 \beta \quad \mbox{and} \quad
 \omega_8=\omega_9=K^2 \beta \lambda \,.
\label{5weights}
\end{equation}
Then $K$ roles as the fugacity per monomer, $\beta$ is a
contact interaction, and $\lambda$ occurs when different loop species
cross as before. We expect the contact interaction to be irrelevant,
and so we set $\beta=1$ in what follows.  The dilute coupled loop
models therefore have partition function
\begin{equation}
Z_{\rm dilute} =
\sum_{{\cal L}} \sum_{{\cal M}} K^{N_{\rm m}} 
Q^{(n_{\cal L}+n_{\cal M})/2}\lambda^{n_X}
\label{Zdilute}
\end{equation}
where $N_{\rm m}$ is the total number of monomers, and the sum is over dilute
loops with the vertices described above. The completely packed model
is recovered in the limit $K \to \infty$.

A critical point in this dilute loop model was discussed at length in
\cite{Floops}.  There several indirect arguments were given as to why
this dilute loop model will have a critical point described precisely
by the {\em same} coset conformal field theory (\ref{coset}) as we
showed above described the critical point at $\lambda=\lambda_c$ in
the completely packed limit. In this section we review these
arguments, use numerics to provide further strong support for them,
and map out the full phase diagram of the dilute loop model.

The arguments presented in \cite{Floops} were the reverse of those
presented above. In \cite{Floops}, critical theories known to describe
the coset models (\ref{coset}) were shown to be closely related to
this dilute loop model. One argument involves an integrable lattice
height model \cite{DJMO,DJKMO} which has a critical point whose
continuum limit is described by the coset (\ref{coset}). It was shown
that in a certain off-critical limit, the height model reduces
precisely to the dilute loop model. An integrable line of couplings
connects this limit to the critical point, so that the corner-transfer
matrix eigenvalues and hence the local height probabilities can be
computed along this line. The configurations which dominate this
computation are precisely those of the dilute loop model. Moreover,
the remaining configurations can be described by a more general dilute
loop model, where multiple strands of each species are allowed on a
link.  The fact that the configurations with multiply occupied links
do not dominate the corner-transfer matrix computation is a strong
sign that at the critical point, allowing or disallowing such
configurations is irrelevant. Thus the dilute loop model discussed
above should still have a critical point described by (\ref{coset}).

Another argument briefly mentioned in \cite{Floops} comes from
studying the $S$ matrix of the integrable field theory at and near
this critical point. A compelling picture of the relation between loop
models and integrable $S$ matrices was presented in
\cite{Zpoly}. There it was argued that one could think of the
worldlines of particles in the integrable $1+1d$ $S$ matrix
description as dilute loops in the corresponding classical 2d lattice
model. This observation was made precise in \cite{Smirnov92}, where it
was shown that the $S$ matrices for the $\Phi_{1,3}$ perturbation of
the minimal models can be expressed in terms of the Temperley-Lieb
generators, which in turn have the loop representation described in
section (\ref{sec:Potts}). For the $S$ matrix ${\cal S}_p$ for the
minimal model of central charge $c=1-6/[(p+1)(p+2)]$,
the weight of closed loops in this representation is
$2\cos(\pi/(p+1))$ (note the shift in the denominator from the weight
of the loops in the completely packed lattice model).
It is very natural to think of these world lines
as domain walls in the 2d classical picture, and this $S$ matrix
indeed describes the loop model in its {\em dilute} phase. Off the
critical point, these domain walls have an effective weight per unit
length.

The coupled loop models can be discussed in the same $S$ matrix
picture.  In fact, it is no more difficult to treat the more general case
\begin{equation}
 \frac{SU(2)_k \times SU(2)_l}{SU(2)_{k+l}}
 \label{cosetkL}
\end{equation}
which will be discussed in more detail in the next section. This
conformal field theory is integrable when perturbed by the
(1,1;adjoint) operator (the analog of $\phi_{1,3}$ in the minimal
models) of dimension $2(1-1/(k+l+2))$. This integrable field theory
describes the lattice height model of \cite{DJMO} near the critical
point, along the line of couplings mentioned above. The $S$ matrices
for this field theory have the interesting property that they
decompose into the tensor product of $S$ matrices of the minimal
models ${\cal S}_{k} \times {\cal S}_{l}$ \cite{Z91}. Each of ${\cal S}_{k}$
and ${\cal S}_{l}$ involves a single Temperley-Lieb algebra, so their
tensor product can be written in terms of tensor products of
Temperley-Lieb generators, namely $1\otimes 1$, $e\otimes f$,
$1\otimes f$, and $e\otimes 1$. The loop representation of these is
exactly that pictured in (\ref{fig:4vertices}) above!  Thus, the
heuristic interpretation of this $S$ matrix is that it describes a
loop model exactly like ours. Since the $S$ matrix is a tensor
product, each particle has structure in both Temperley-Lieb algebras,
so its worldlines must be described as doubled lines, and the lines of
the different copies can cross at a vertex.

Still more evidence that the dilute loop model has a critical point
comes from another solvable model \cite{Grimm} closely related to
our dilute loop model. The models in \cite{Grimm}
are based on the ``dilute BMW'' algebra, generalizing those of \cite{JMO}
with weights (\ref{Ru}) to allow heights on adjacent sites to be the
same. In the graphical representation, this amounts to relaxing the
complete-packing requirement. As we have described above in section
\ref{sec:critpoint}, the $SO(4)$ BMW algebra decomposes into the
direct product of two Temperley-Lieb algebras. Thus the dilute $SO(4)$
BMW algebra in its loop representations describes the same dilute loop
model. To make it integrable, one must tune various couplings,
including that of the vertex where the loops go straight across.
Unfortunately, the corner transfer matrix computation of these models
does not seem to have been done, so the CFT describing this critical
point has not yet been identified. However, by using the inversion
relation method, one can easily compute a critical exponent at each of
the two critical points in this model. At the critical point where the
weights favor dilute loops, the critical exponent found corresponds to
an operator of dimension $3/(k+2)$, which indeed is present in the
coset model (\ref{coset}). Unfortunately, at the isotropic point some
Boltzmann weights are negative (although the corresponding $1+1$
dimensional quantum Hamiltonian is hermitian), so it is not
possible to make a precise comparison with our dilute loop model. 

All this is compelling evidence that the dilute loop has a critical
point with the same conformal field theory description as that of the
completely packed model at $\lambda=\lambda_c$. Minor modifications in
the definition of the the loop model, such as allowing multiple strands
on a link, or loops to go straight through a vertex, do not change the
conclusion.

There are no marginal operators in this coset conformal field theory
describing the dilute and completely packed critical point. Thus there
should be a flow in the loop model connecting the two points.

Several compelling pieces of evidence imply that the completely packed
critical point is stable, i.e., including dilution is an irrelevant
perturbation. It has long been known that dilution is irrelevant in
the context of a single Potts model. An initial Coulomb gas-type (and
thus non-rigorous) argument \cite{Nienhuis_dilution} was subsequently
placed on firm ground through the integrability analysis of the most
general model of self-avoiding loops defined on the square lattice
that is compatible with an O($n$) symmetry
\cite{Nienhuis89,Nienhuis90}. For a given value of $n \in (-2,2)$,
this model admits five distinct integrable points with isotropic
weights at the vertices. One is the completely packed model which is
equivalent to the $Q=n^2$ state Potts model. Two other solutions give
the so-called dense and dilute solutions for the O($n$) model, and
fall in the same universality class \cite{Warnaar93a,Warnaar93b} as
those found previously \cite{Nienhuis82} for the honeycomb lattice
\cite{Batchelor88}. Crucially, the dense solution is in the same
universality class as the Potts model solution, and so we learn that a
deviation from complete packing is an irrelevant perturbation. [The
remaining two solutions are multicritical extensions of the dense and
dilute solutions.]

Returning to our case of two coupled models, we thus have that at the
decoupled critical points at $\lambda=\pm 1$, the flow in $K$ is
toward the completely packed line ($K=\infty$). It is natural to
assume that this persists for all $\lambda$, so that in particular
dilution is irrelevant both throughout the critical phase discussed in
section \ref{sec:critphase} and near the critical point discussed in
section \ref{sec:critpoint}.

To complete this picture we turn to numerical transfer matrix calculations.
We have been able to study the dilute model (\ref{Zdilute}) for periodic
strips of width up to $L=14$.%
\footnote{Note that the exclusion of straight-going
vertices from Fig.~\ref{fig:5vertices} is very convenient, since it
reduces the dimension of the Hilbert space underlying the transfer matrix
construction. Indeed, each element of the space is a pair (one for each
loop species) of non-perfect matchings of $L$ points, subject to the usual
constraint of planarity with respect to each loop species. In the absence of
straight-going vertices, each element in the matching is a pair of points on
sublattices of {\em opposite} parity. With straight-going vertices allowed,
there is no such parity constraint.}
\begin{figure}
\begin{center}
\includegraphics[width=9.4cm,angle=270]{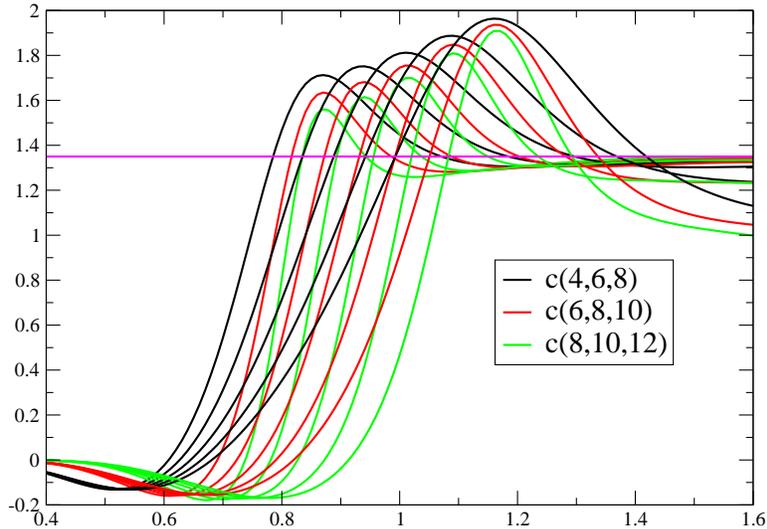}
\end{center}
\caption{Central charge $c$ as a function of $K$ in the dilute model.
  The parameter $k=3$, corresponding to $c=\frac{27}{20}=1.35$ (shown
  as a pink horizontal line) at
  $\lambda_c=-\sqrt{2}\sin(\frac{\pi}{20}) \simeq -0.2212$.  Curves of
  the same color correspond to several different values of
  $\lambda=-0.1, -0.2, -0.3, -0.4, -0.5$. As $\lambda$ decreases the
  curves are shifted towards the right.}
 \label{fig:cdiluteB5}
\end{figure}
Results for the central charge $c$ as a function of $K$, but fixed
$\lambda$, are shown in Fig.~\ref{fig:cdiluteB5}, still for the case
$k=3$. Similar results hold true for other parameter values $k>2$. The
various curves correspond to different choices of $\lambda$. The
reader may train his/her eye to view this as a three-dimensional plot
of $c$ as a function of the two parameters $(K,\lambda)$. Applying
Zamolodchikov's c-theorem---which states that RG flows go
downhill---reveals that for a large enough $K$ the flow is towards
$(K,\lambda) = (\infty,\lambda_c)$, i.e., the critical point of the
completely packed model. By ``large enough'' is meant that
$K>K_c(\lambda)$, where $K_c(\lambda)$ is the locus of the global
maximum in $c$. On the other side of the phase boundary, $K <
K_c(\lambda)$, the flow is towards a non-critical phase, e.g.\ the
trivial fixed point where all loops are absent, or the
$(K,\lambda)=(\infty,0)$ single Potts model point $S$. It is readily
verified that dilution is also irrelevant throughout the critical
phase. Simulations on the line $\lambda=0$ show that for $k>2$ dilution
decreases the effective $c$, so the flow is towards $K=0$.

\begin{figure}
\begin{center}
\includegraphics[width=8cm,angle=0]{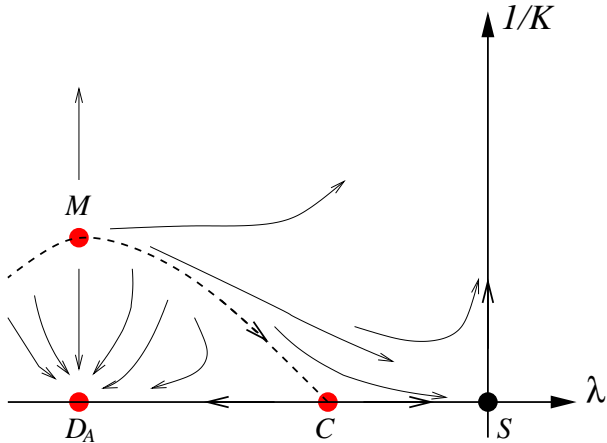}
\end{center}
\caption{Schematic phase diagram for the dilute model with $k>2$.}
 \label{fig:phase_dilute}
\end{figure}

We therefore arrive at the schematic phase diagram depicted in
Fig.~\ref{fig:phase_dilute}. 
The RG flows are shown as arrows.  
A continuous first-order phase transition occurs
along the dashed line. The points $S$ (at $\lambda=0$), $C$
(at $\lambda=\lambda_c$) and $D_A$ (at $\lambda=-1$) are the same as in
the completely packed loop model (see Fig.~\ref{fig:phase_dense}).
The point $M$ is a putative multicritical point, unstable in both
parameters. For this precise lattice model, we do not know if there is
an dilute integrable point like in the lattice models in the same
universality class discussed above. If there is, it would be somewhere
along the dashed line. 
Note that there should be an RG separatrix joining $M$ and $S$, as shown.

We have not been able to identify $M$ exactly using our numerical
scheme. Because of the $\lambda \mapsto 1/\lambda$ duality, it is
natural to assume that $M$ is located on the decoupling line
$\lambda=-1$, but the interference with the first-order transition
line impedes the numerical verification of this expectation.
Also, determining where the various phase boundaries end, or coalesce,
is not an easy matter. We have however verified that the dilute critical
phase extending from $C$ is also stable towards a perturbation in the
hitherto ignored parameter $\beta$ of (\ref{5weights}).

If we generalize the coupled dilute model to include all the
interactions present in the most general O($n$) symmetric loop model
referred to above \cite{Nienhuis89,Nienhuis90} we certainly have three
candidate multicritical points along the decoupling line $\lambda=-1$
(viz. dilute, multicritical dense, and multicritical dilute). Our
constraints on the parameter space are however likely to make at least
two of these points disappear. On the other hand, moving away from
$\lambda=-1$ could in principle lead to a different kind of
multicriticality.

\section{The self-dual critical point for $Q_1\ne Q_2$}
\label{sec:differentQ}

At the beginning of section \ref{sec:dilute}, we reviewed the
arguments of \cite{Floops} indicating that the dilute coupled loop
models had a critical point. Both the $S$ matrix and the corner
transfer matrix arguments apply to a more general class of coset
models, those built on the cosets (\ref{cosetkL}).
The central charge of this critical point is 
\begin{equation}
 c=\frac{3 k l (4 + k + l)}{(2 + k) (2 + l) (2 + k + l)}
 \label{central_kl}
\end{equation}
In the loop language the parameters $k$ and $l$ parameterize the weights
for the two different species of loops:
$$\sqrt{Q_k}=2\cos \left( \frac{\pi}{k+2} \right) \quad \mbox{and} \quad
  \sqrt{Q_l}=2\cos \left( \frac{\pi}{l+2} \right) \,.$$
These correspond simply to replacing (\ref{Zdilute}) by
\begin{equation}
 Z=\sum_{\cal L} \sum_{\cal M} K^{N_{\rm m}} 
 Q_k^{n_{\cal L}/2}Q_l^{n_{\cal M}/2}\lambda^{n_X}
\label{Zdilute2}
\end{equation}
so that we are now coupling a $Q_k$-state and a $Q_l$-state Potts model.

It is thus natural to assume that this loop model with $k \ne l$ will
have the same phase diagram as that established above for $k=l$. In
particular, there should be a non-trivial critical point in the
completely packed model at some negative value of $\lambda$, described
in the continuum limit by the coset conformal field theory
(\ref{cosetkL}).

We have checked this numerically for several integer values of the parameters
$k$ and $l$. Based on the maxima of the central charge
(cf.~Fig.~\ref{fig:centralB5}), we find in particular
\begin{eqnarray}
 c = 1.4675(4) \mbox{ at } \lambda_c = -0.2908(7) \quad
 \mbox{for } (k,l)=(3,4) \nonumber \\
 c = 1.5431(3) \mbox{ at } \lambda_c = -0.3376(2) \quad
 \mbox{for } (k,l)=(3,5)
\end{eqnarray}
This agrees nicely with (\ref{central_kl}) which predicts
$c = \frac{22}{15}  = 1.46667$ for $(k,l)=(3,4)$ and
$c = \frac{54}{35}  = 1.54286$ for $(k,l)=(3,5)$. 
We have unfortunately not been able to provide a convincing
conjecture for $\lambda_c$ to generalize the exact result (\ref{lamc}).

\bigskip

The research of PF has been supported by the NSF under grants
DMR/MSPA-0704666 and DMR-0412956, and by an EPSRC grant EP/F008880/1.
The research of JLJ has been supported by the European Community
Network ENRAGE (grant MRTN-CT-2004-005616) and by the Agence Nationale
de la Recherche (grant ANR-06-BLAN-0124-03). The authors thank the
Isaac Newton Institute for Mathematical Sciences, where part of this
work was done, for hospitality.

\end{document}